\documentclass[reprint, superscriptaddress, amsmath,amssymb, aps, pra, floatfix]{revtex4-1}
\usepackage[T1]{fontenc} 
\usepackage{mathptmx}
\usepackage{graphicx}
\usepackage{dcolumn}
\usepackage{bm}
\usepackage{multirow}
\usepackage{makecell}
\usepackage[colorlinks, breaklinks, citecolor=blue,linkcolor=black,urlcolor=black]{hyperref}

\begin{document}

\title{Three-dimensional Magneto-optical Trapping of Barium Monofluoride}

\author{Zixuan Zeng}
\author{Shuhua Deng}
\author{Shoukang Yang}
\affiliation{Zhejiang Key Laboratory of Micro-nano Quantum Chips and Quantum Control, School of Physics, and State Key Laboratory for Extreme Photonics and Instrumentation, Zhejiang University, Hangzhou 310027, China
}
\author{Bo Yan}
\email{yanbohang@zju.edu.cn}
\affiliation{%
Zhejiang Key Laboratory of Micro-nano Quantum Chips and Quantum Control, School of Physics, and State Key Laboratory for Extreme Photonics and Instrumentation, Zhejiang University, Hangzhou 310027, China
}%
\affiliation{College of Optical Science and Engineering, Zhejiang University, Hangzhou 310027, China
}

\date{\today}

\begin{abstract}
As a heavy molecule, barium monofluoride (BaF) presents itself as a promising candidate for measuring permanent electric dipole moment. The precision of such measurements can be significantly enhanced by utilizing a cold molecular sample. Here we report the realization of three-dimensional magneto-optical trapping (MOT) of BaF molecules. Through the repumping of all the vibrational states up to $v=3$, and rotational states up to $N=3$, we effectively close the transition to a leakage level lower than $10^{-5}$. This approach enables molecules to scatter a sufficient number of photons required for laser cooling and trapping. By employing a technique that involves chirping the slowing laser frequency, BaF molecules are decelerated to near-zero velocity, resulting in the capture of approximately $3\times 10^3$ molecules in a dual-frequency MOT setup. Our findings represent a significant step towards the realization of ultracold BaF molecules and the conduct of precision measurements with cold molecules.

\end{abstract}

\maketitle
Molecules have important applications in the pursuit of new physics through precise measurements, such as the electric dipole moment (eEDM) measurement and parity violation \cite{Carr2009, Safronova2018, Langen2024}. Heavy molecules, such as ThO \cite{Meyer2008}, TlF \cite{Cho1989, Grasdijk2021}, YbF \cite{Hudson2011}, RaF \cite{Isaev2010}, BaF \cite{Aggarwal2018} and certain polyatomic molecules \cite{Kozyryev2017a, Jadbabaie2023} exhibit significant enhancements of the effective electric field within a molecule, making them prime candidates for eEDM measurement. Currently, the most sensitive eEDM measurements are conducted using molecular beams \cite{Andreev2018} or molecular ions \cite{Roussy2023}.

Drawing from advancements in atomic physics, where precision measurements benefit greatly from cold atoms \cite{Bothwell2022, Overstreet2022}, cold molecules could potentially play a similar role in precision measurements. There are already some proposals to use cold molecules, such as  YbF \cite{Tarbutt2013}, BaF \cite{Aggarwal2018}, RaF \cite{Ruiz2020}, and some polyatomic molecules \cite{Kozyryev2017a}, to refine eEDM measurements. Among them, the precision spectroscopy has been obtained for RaF, along with a proposed laser cooling scheme \cite{Udrescu2024}; SrOH, YbOH, YbF, and BaF have been successfully laser-cooled in the transverse direction \cite{Kozyryev2017, Augenbraun2020, Lim2018, Zhang2022, Rockenhaeuser2024}.  However, the realization of magneto-optical trapping (MOT) for these heavy molecules is still challenging.

On the other hand, laser cooling and trapping of molecules have witnessed remarkable achievements in recent decades. Approximately ten molecules have been successfully laser-cooled \cite{Shuman2010, Hummon2013, Zhelyazkova_2014, Kozyryev2017,Lim2018, Baum2020,Augenbraun2020, Mitra2020,McNally2020,Zhang2022,  Rockenhaeuser2024, Dai2024}. However,  only four of them, SrF \cite{Barry2014}, CaF \cite{Truppe2017, Li2024}, YO \cite{Collopy2018, Burau2023}, and CaOH \cite{Vilas2022} have been reported to be trapped in the MOT. Some have been further trapped using magnetic traps \cite{McCarron2018,Williams2018}, dipole traps \cite{Cheuk2018,Anderegg2018,Hallas2023, Jorapur2024,Burau2024}, optical lattices \cite{Wu2021} and optical tweezers \cite{Anderegg2019}. 
Notably, the combination of molecular MOT and optical tweezer overcomes the limitation of low density and reveals interesting physics such as dipole exchange interaction and entanglement \cite{Bao2023, Holland2023}.

These significant advancements have led to a flourishing era in the field of laser cooling and trapping of molecules. However, the task of realizing a molecular MOT for a new species remains highly challenging. First, it is essential to identify all the relevant transitions with high resolution. Some transitions, such as the hyperfine splitting of the excited states \cite{Bu2022, Denis2022, Rockenhaeuser2023}, may never have been measured before. Second, all the important leakages with ratios higher than $10^{-5}$, such as the higher vibrational states and some rotational states, must be found and recycled. For some complex molecules \cite{Zhu2022, Zeng2023}, such a task is quite challenging. 

\begin{figure}[]
\centering
\includegraphics[width=0.45\textwidth]{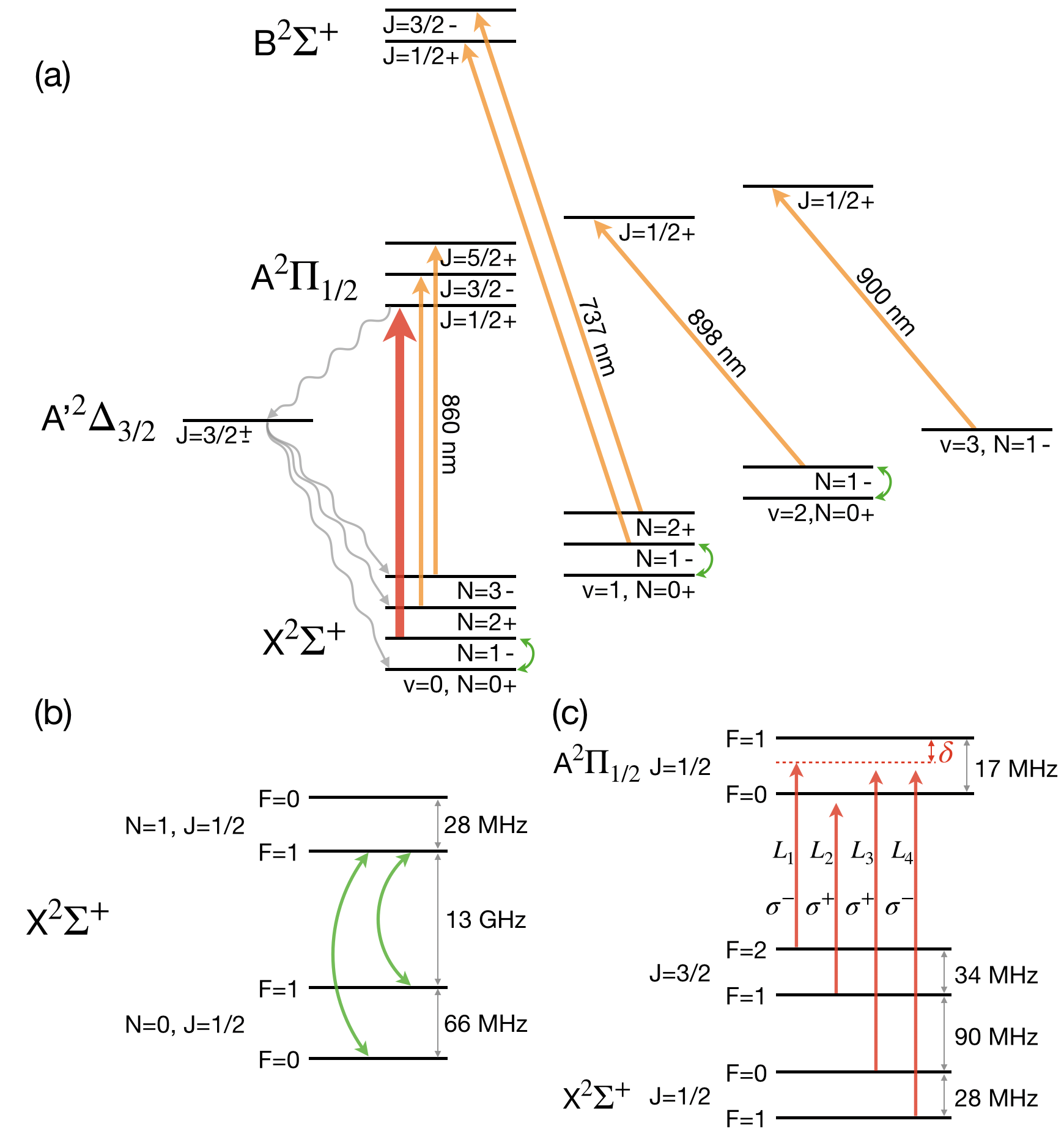}
\caption{(color online) \label{fig:levels}
Energy levels involved in the BaF MOT experiment. (a) shows the transitions for the main cooling lasers (red arrow line) and the repump lasers (orange arrow lines). The green arrow lines indicate the microwave remixing. (b) shows the microwave remixing scheme in the $v=0$ manifold. The two hyperfine states in $|N=0, J=1/2\rangle$ are coupled with $|N=1, J=1/2, F=1\rangle$ state with two 13~GHz microwaves.  (c) shows the hyperfine structures of the main cooling transition. The main cooling laser has four components $(L_1-L_4)$ with frequency separations of $\{0, 6, 122, 148\}$~MHz, The detuning $\delta$ is defined with $L_1$ accordingly as shown in the plot. In our experiment, it is red detuned and $\delta=4$~MHz.
}
\end{figure}

Here, building on our earlier work with BaF molecules on the high-precision spectroscopy \cite{Bu2017, Bu2022}, laser deflection \cite{Chen2017} and laser cooling \cite{Zhang2022}, we present the successful realization of a 3D MOT for BaF molecule.
The relevant laser cooling and repump transitions are illustrated in Fig. \ref{fig:levels}. The main cooling transition is $|X^2\Sigma, v=0, N=1\rangle\to|A^2\Pi_{1/2},v'=0,J'=1/2,+\rangle$, with a braching ratio of approximately $95\%$. Notably, the most significant leakage occurs through the transition $|A^2\Pi_{1/2},v'=0\rangle\to |X^2\Sigma,v=1\rangle$, with a branching ratio of about $4.7\%$. To address this, a 737~nm laser is employed to repump these $v=1$ states back through the $B^2\Sigma$ state \cite{Zhang2022}. Additionally, the leakage ratios to $v=2$ and $v=3$ are approximately $2\times 10^{-3}$ and $6\times 10^{-5}$, respectively, through both the excited $A^2\Pi_{1/2}$ and $B^2\Sigma$ states \cite{Chen_2016,Albrecht2020,Hao2019}. To mitigate these leakages, we utilize 898~nm and 900~nm lasers for repumping, as depicted in Fig. \ref{fig:levels}(a). In our current experiment, the leakage ratio to even higher vibrational states should be less than $10^{-5}$ and can thus be neglected.

One of the difficulties in realizing the BaF MOT lies in the fact that there is a low-lying metastable $A'^2\Delta_{3/2}$ state, akin to YO molecules \cite{Yeo_2015}. The branching ratio of $|A^2\Pi_{1/2}\rangle\to|A'^2\Delta_{3/2}\rangle$ occurs at $10^{-4}$ level, where molecules in the excited  $A^2\Pi_{1/2}$ state decay back to the ground states through this intermediate $\Delta$ state via a two-photon process. This results in a parity change from the initial state. As such, molecules decay to the $N = 0$ and 2 ground rotational states instead of $N = 1$ states. Moreover, the nearly degenerate states with different parity in $A'^2\Delta_{3/2}$ ($|J'=3/2,+\rangle$ and $|J'=3/2,-\rangle$) are susceptible to mixing with any residual electric field, leading to leakages to $N=1$ and 3 states. Theoretical results of branching ratios \cite{Chen_2016, Albrecht2020, Hao2019} indicate that the ground rotational states with leakage ratios higher than $10^{-5}$ include $|v=0, N=0,2,3\rangle$ and~$|v=1, N=0,2\rangle$. Consequently, all these dark states necessitate careful attention and repumping.

To address the $N=0$ states, we employ microwaves to couple $N=0$ and $N=1$ states as illustrated in Fig. \ref{fig:levels}(a).  Figure \ref{fig:levels}(b) shows the hyperfine structure in the $v=0$ manifold. Both $|N=0,F=0, 1\rangle$ states are mixed with $|N=1,F=1\rangle$. The rotational transition frequencies slightly differ for different vibrational states. Specifically, we utilize three microwaves with frequencies: $\omega_1=12.906$~GHz, $\omega_2=12.838$~GHz, and $\omega_3=12.770$~GHz, which cover these rotational transitions in $v=0,1,2$ manifolds (see Table \ref{Table1})..

\begin{figure}[]
\centering
\includegraphics[width=0.49\textwidth]{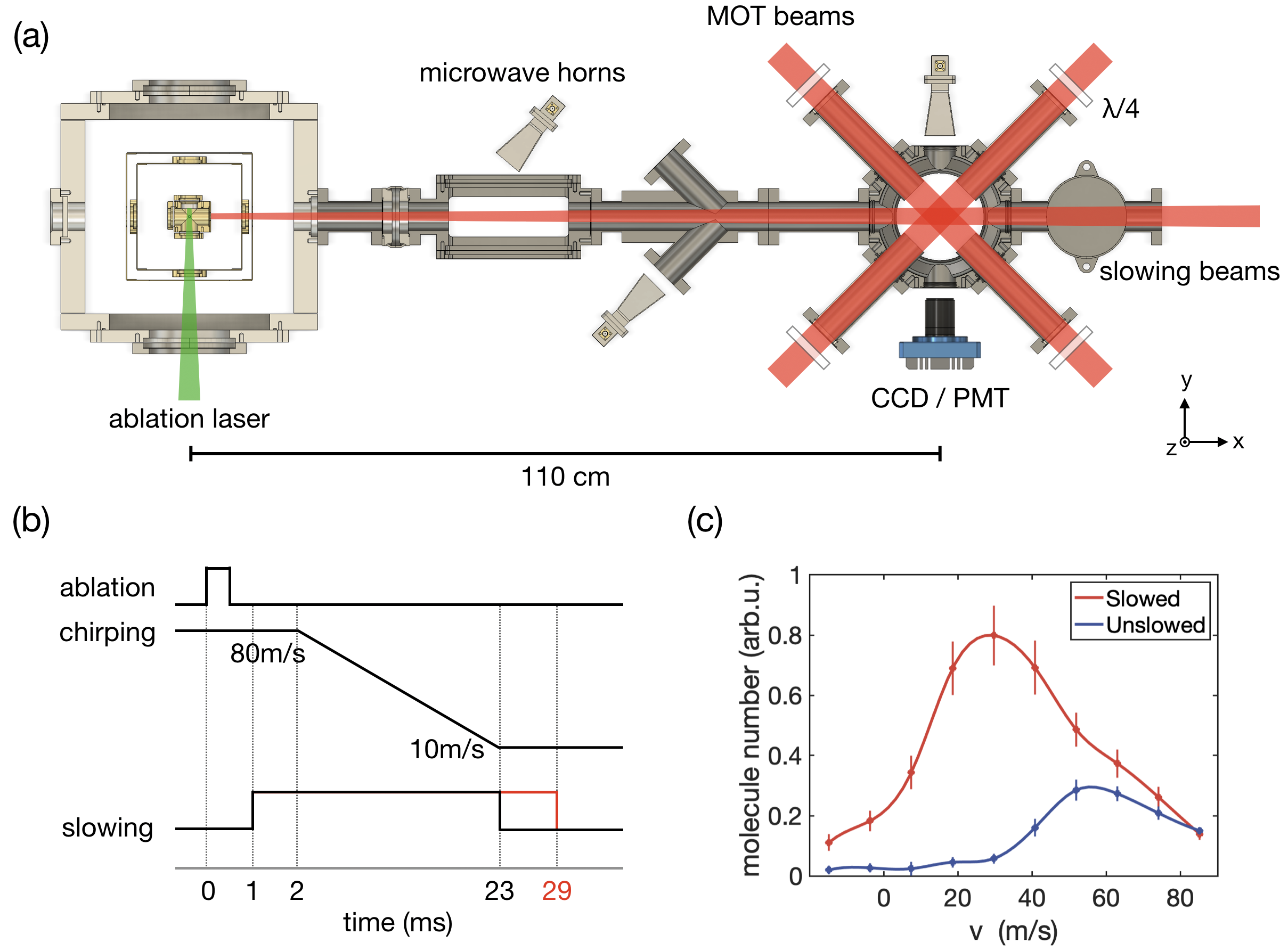}
\caption{(color online) \label{fig:slowing}
(a) Top view of the experiment apparatus. The molecule created in the buffer gas cell travels 110~cm to the MOT region and gets detected. Three microwave horns are used to remix the $N=0$ states. (b) The time sequence used for frequency-chirped slowing. There is a 6 ms extra slowing in the MOT experiment, as highlighted with the red lines. (c) The velocity distribution of molecules with and without slowing during 24~ms to 38~ms after laser ablation.
}
\end{figure}

\begin{table}[!ht]
\caption{Relavent transitions and microwaves used to recycle the $N=0$ states in the ground state.}\label{Table1}
\begin{tabular}{|c|c|c|c|c|} \hline 
$v$ & \makecell[c]{State in \\ $N=0$} & \makecell[c]{State in \\ $N=1$} & \makecell[c]{Trans. Freq. \\ $\omega$ (GHz)}& Mixed by \\ \hline
\multirow{2}{*}{$v=0$} & $J=1/2,F=0$ & $J=1/2,F=1$ & 12.904 & $\omega_1$ \\  \cline{2-5}
	~ &	$J=1/2,F=1$ &	$J=1/2,F=1$ & 12.838  & $\omega_2$\\ 
\hline
 \multirow{2}{*}{$v=1$} & $J=1/2,F=0$ & $J=1/2,F=1$ &	12.834  & $\omega_2$\\  \cline{2-5}
 ~ & $J=1/2,F=1$ & $J=1/2,F=1$ & 12.768   & $\omega_3$\\ 
\hline
	\multirow{2}{*}{$v=2$} &	$J=1/2,F=0$ &	$J=1/2,F=1$ &  12.849  & $\omega_2$\\  \cline{2-5}
 ~ & $J=1/2,F=1$ & $J=3/2,F=2$ &	12.765  & $\omega_3$\\ \hline
\end{tabular}
\end{table}

To address the $N=2,~3$ states in $v=0, 1, 2$ manifolds, we employ lasers for repumping as illustrated in Fig. \ref{fig:levels}(a). They are mainly generated by the sideband of fiber EOMs, and the powers are about a few mW. This optical-microwave mixed repump scheme effectively reduces the number of ground states, thereby increasing the maximum scattering rate  \cite{Collopy2018} to $\Gamma/14$ in our case, where $\Gamma=2\pi\times 2.8~$MHz is the natural linewidth of the excited $A^2\Pi_{1/2}$ state. With these vibrational and rotational repumpings, we close the main cooling transition to a leakage level lower than $10^{-5}$.

Figure \ref{fig:slowing} (a) shows the top view of the experimental setup. BaF molecules are produced by ablation of BaF$_2$ targets, and are cooled in a single-stage cryogenic buffer gas cell at 4 K, and exit through a 4 mm aperture. The buffer-gas-cooled molecular beam has a peak forward velocity of approximately 90~m/s with a Helium flow rate of 0.2 sccm. The distance from the cell to the MOT regime is 110 cm. Along the chamber, three microwave horns are positioned as shown in Fig. \ref{fig:slowing}(a) to introduce the microwaves. The vacuum pressure in the MOT chamber is $7\times10^{-7}$ Pa and increases to $4\times10^{-6}$ Pa when the Helium gas flow is on.

The slowing beam consists of the main slowing laser and all the repump lasers, with a beam waist of 11 mm. It propagates against the molecular beam and is slightly focused near the cell position. The main slowing laser (860~nm) has a total power of 140~mW and contains two frequency components separated by $134~$MHz. Additionally, it is modulated by a 12.4~MHz EOM to broaden the spectrum up to the 2nd sidebands. To compensate for the Doppler shift during the slowing process, the slowing lasers are frequency chirped as depicted in Fig. \ref{fig:slowing}(b). Correspondingly, the slowed velocity is ramped from 80~m/s to 10 m/s in 21~ms. All the other repumping lasers are modulated with a 38 MHz EOM for hyperfine splitting and a 5.4 MHz EOM for spectrum broadening, resulting in a white light with a frequency span of approximately 200 MHz. During the slowing process, there is also frequency chirping of the repumping lasers.  

\begin{figure}[]
\centering
\includegraphics[width=0.42\textwidth]{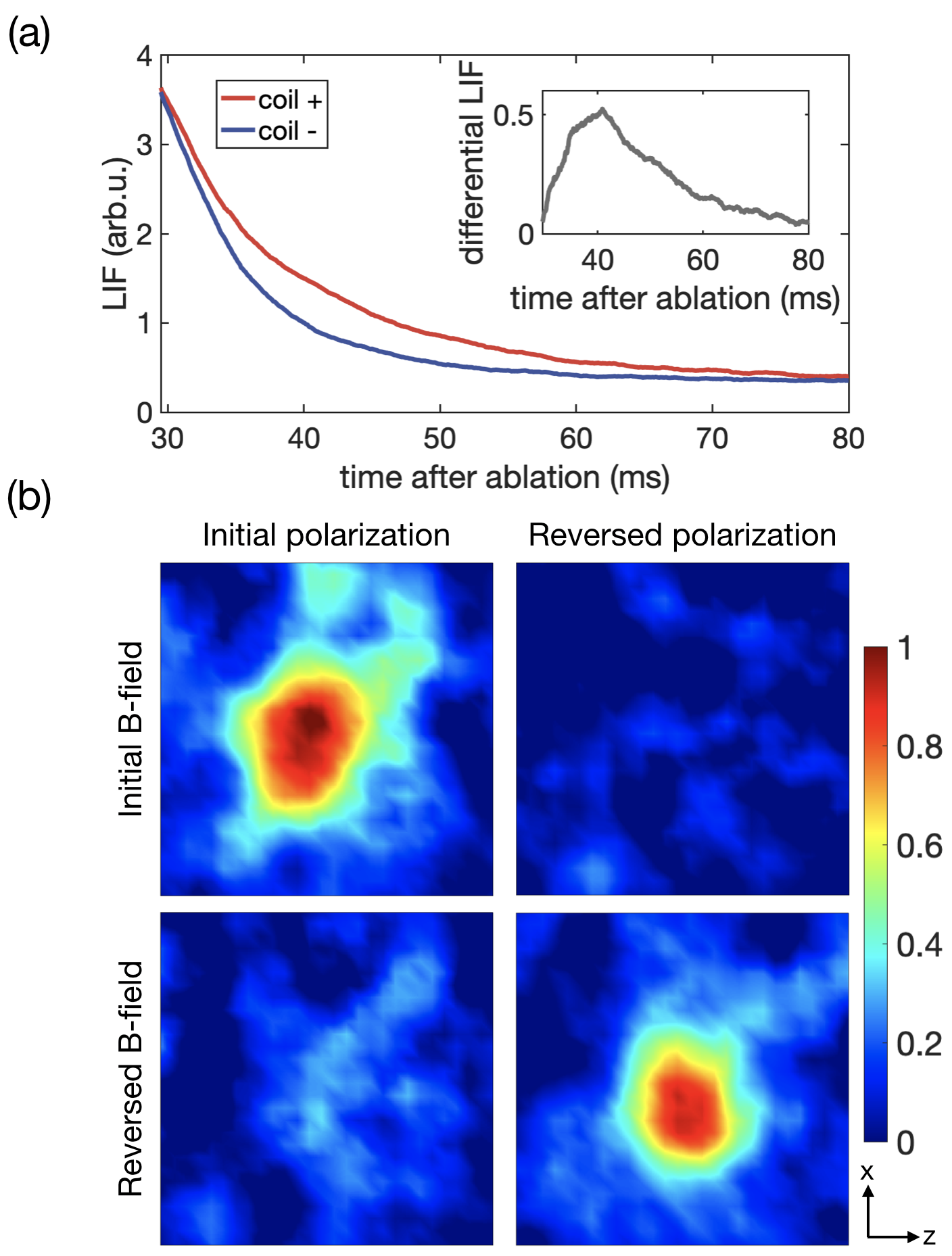}
\caption{(color online) \label{fig:LIF}
 (a) the LIF signal of the molecules at the MOT regime. For fixed laser polarizations, we flip the current direction in the magnetic coils to get two LIF signals. Insert shows the subtracted LIF time trace of MOT loading and decay acquired from the PMT. (b) The BaF MOT signals at different conditions. The four pictures are images taken from 60~ms to 80~ms with different polarizations and signs of $dBz/dz$. Each plot presents an average of 240 times and has a size of 1~cm $\times$ 1~cm.
}
\end{figure}

The MOT beam, comprising the main cooling laser and $v=1$ and $v=2$ repump lasers, has a beam size of 14 mm. The main cooling laser has a power of 20 mW and consists of four components ($L_1-L_4$) with frequency separations of $\{0, 6, 122, 148\}$MHz. To create the dual-frequency MOT \cite{Tarbutt2015}, two components ($L_1$ and $L_4$) have a polarization of $\sigma-$, while the other two ($L_2$ and $L_3$) are $\sigma+$ polarization, as illustrated in Fig. \ref{fig:levels} (c). 
The repump powers are 60 mW and 10 mW for $v=1$ and $v=2$, respectively.  

We utilize a PMT to detect the light-induced fluorescence (LIF) signal of molecules. To minimize the impact of stray light from the MOT beams, a 712~$nm$ bandpass filter is placed in front of the PMT. The 712~nm fluorescence is emitted due to the transition of $|B^2\Sigma, v=0\rangle\to|X^2\Sigma, v=0\rangle$, which is initiated by the $v=1$ repump laser (737~nm). With a branching ratio of $4.7\%$ for the transition $|A^2\Pi_{1/2}, v=0\rangle\to|X^2\Sigma,v=1\rangle$, this 712~nm LIF signal is approximately 20 times weaker than the 860~nm fluorescence. Nevertheless, this setup effectively eliminates all the stray light, ultimately improving the signal-to-noise ratio in our current experiment. 

Using the LIF signal, we can extract the velocity distribution of molecules at the detection region with a Doppler-sensitive detection, with the probe beam being sent 45 degrees across the molecule beam. Figure \ref{fig:slowing}(c) illustrates the typical molecule velocity distribution with and without the slowing process over a duration of 24~ms to 38~ms. Since the detection occurs 24~ms after the laser ablation, only a small fraction of slower molecules (around 60~m/s) is observable without slowing, and almost no signal for molecules with velocities lower than 30~m/s. However, with the frequency-chirped slowing exhibited in Fig. \ref{fig:slowing}(b), there is a significant enhancement in the number of molecules at low velocities. In this way, we have effectively slowed the buffer-gas-cooled BaF molecules to near-zero velocity,  making a promising starting point to search for a molecular MOT signal. 


\begin{figure}[]
\centering
\includegraphics[width=0.4\textwidth]{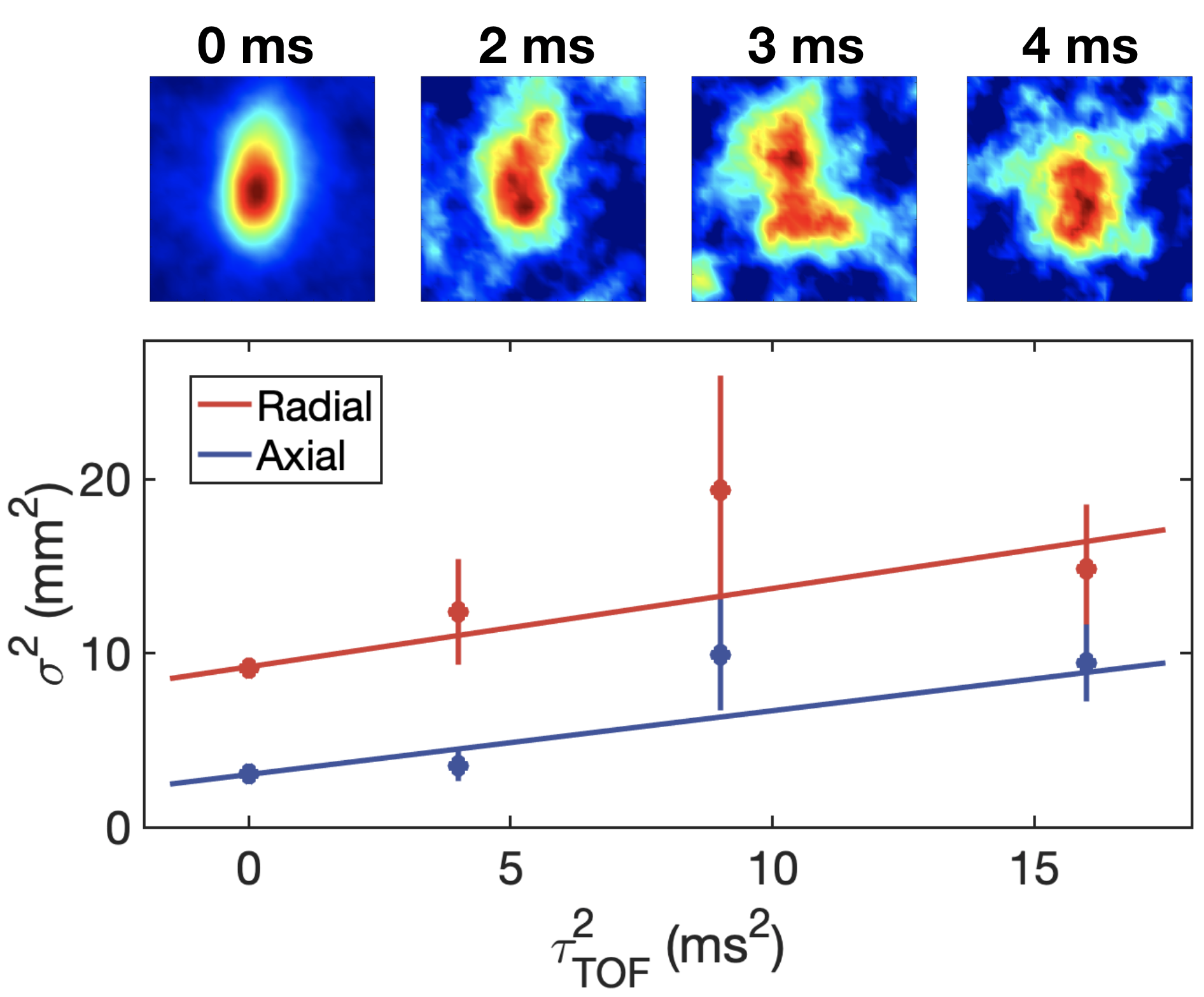}
\caption{(color online) \label{fig:temp}
The temperature measurement of the BaF molecule. The above figures show the molecular MOT images at different expansion time $\tau_\text{TOF}$. Each plot presents an average of 400 times and has a size of $1.5~cm\times 1.5~cm$. The below figure shows the dependences of $\sigma^2$ versus $\tau_\text{TOF}^2$ and is fitted with a linear function, which gives the temperature as $T_\text{radial}=9(4)~mK$ and $T_\text{axial}=7(3)~mK$ in both radial and axial directions. }
\end{figure}

The magnetic field in the MOT is generated by a pair of anti-Helmholz coils, with a magnetic gradient set to $dB_z/dz$=8~G/cm.  Notably, even in the absence of a MOT, the LIF signal is still detectable as slow molecules pass through the MOT region and are illuminated by the MOT beam.  To minimize the impact of these flying molecules, we alternate the current direction of the MOT coil at each cycle, referred to as the "coil $+$" and "coil $-$" conditions. The laser polarizations illustrated in Fig. \ref{fig:levels} (c) are defined under the "coil $+$" condition, allowing for the magneto-optical trapping of molecules. No MOT is expected under the "coil $-$" condition.  Figure \ref{fig:LIF} (a) displays LIF time traces in the MOT region, with the red and blue lines representing two datasets collected under different magnetic field conditions.  The insert plot shows the differential signal for these two datasets, providing strong evidence of a BaF MOT. The MOT begins loading at 29~ms and has a lifetime of approximately 20~ms.  It is worth mentioning that extending the slowing process for a few more milliseconds is preferable for MOT loading. All the MOT data in this paper is taken with 6~ms extra holding time as shown in Fig. \ref{fig:slowing} (b).

We also employ a CCD to capture images of the molecular MOT cloud. Figure \ref{fig:LIF} (b) displays images taken with different polarizations and signs of $dB_z/dz$. The image acquisition begins at 60~ms after ablation and lasts for 20~ms. Under the appropriate polarization and magnetic field gradient direction, a MOT signal becomes apparent. When we switch both the polarization and the magnetic field direction, the MOT signal persists, as evidenced by the first and fourth quadrants. If only one of the two conditions is switched, no MOT signal is observed, as depicted in the second and third quadrants. This result provides further evidence of the existence of a molecular MOT. We estimate the molecule number of the BaF MOT is approximately $3\times 10^3$.

To measure the MOT temperature, we switch off the MOT beam at 42~ms, allowing molecules to expand for a certain time $\tau_\text{TOF}$, then we turn on the MOT beam again for 2~ms to image molecules. Figure \ref{fig:temp} shows such measurement. The MOT image is obtained by subtracting the ``coil $+$'' and ``coil $-$'' conditions. We then fit the images with the Gaussian function and extracted the fitted size $\sigma$. By fitting $\sigma^2$ versus $\tau_\text{TOF}^2$ as a linear function, we extract the temperature as $T_\text{radial}$=9(4)~mK in the radial direction and $T_\text{axial}$=7(3)~mK in the axial direction.
Using the equipartition theorem, we can estimate the MOT trap frequencies as $f_\text{radial}=35~$Hz and $f_\text{axial}=55~$Hz.


To conclude, we have successfully achieved a 3D-MOT of BaF molecules, which presents a significant step towards the realization of ultracold BaF molecules and opens up possibilities for eEDM measurement with cold molecules. Due to the presence of a low-lying $\Delta$ state, leakage states may possess different parity from the initial state, necessitating rotational repumping. To address this, we employ repump lasers and microwaves to recycle these dark states of $N=0-3$ in $v=0-3$ manifold. In our current experiment, we utilize a total of 8 lasers, 3 fiber EOMs, and 3 microwaves to recycle those rotational and vibrational dark states. Fortunately, the relevant transitions of the BaF molecule occur at wavelengths around 700~nm - 900~nm, falling within the optimal range for the diode lasers. Consequently, we can utilize homemade external cavity diode lasers (ECDLs) to cover all the required lasers. This repump scheme also makes molecules in the low-lying rotational states ($N=0-3$), not just $N=1$, useful in laser cooling, resulting in a larger number of cooled and trapped molecules.

Currently, we continue to utilize microwaves to recycle the $N=0$ states by remixing them to $N=1$ states. However, in the future, we aim to implement a laser to optical pump $|X^2\Sigma, v, N=0\rangle \to |A'^2\Delta, v'=v, J'=3/2\rangle$ with a wavelength of 933~nm. In $\Delta$ states, the different parity states are nearly degenerate, allowing for an easy intermix of parity with the electric field and recycling back to the $N=1$ manifold.  This approach will reduce the number of ground states involved in laser cooling and consequently increase the declaration force.

We thank Prof. Jun Ye and Prof. Shiqian Ding for helpful discussions and Justin Johnathan Burau for careful reading of our manuscript. We acknowledge the support from the National Natural Science Foundation of China under Grant No. U21A20437 and No. 12074337, the National Key Research and Development Program of China under Grant No. 2023YFA1406703 and No. 2022YFA1404203,  Natural Science Foundation of Zhejiang Province under Grant No. LR21A040002, and the Fundamental Research Funds for the Central Universities under Grant No. 226-2023-00131.

\bibliographystyle{apsrev4-1}
\bibliography{MOT_reference}

\end{document}